\documentclass[12pt]{iopart}

\usepackage[dvipdfmx]{graphicx}

\usepackage[dvipdfm]{hyperref}
\usepackage{amssymb}

\usepackage{cite}

\usepackage[final]{changes}

\newcommand{\subtxt}[1]{\mbox{\scriptsize #1}}

\newcommand{\PoP}[0]{Phys. Plasmas}
\newcommand{\NF}[0]{Nucl. Fusion}
\newcommand{\JPP}[0]{J. Plasma Phys.}

\begin{document}
%\noindent
%This is an author-created, un-copyedited version of an article originally published in Plasma Physics and Controlled Fusion.  IOP Publishing Ltd is not responsible for any errors or omissions in this version of the manuscript or any version derived from it.  The Version of Record is available online at \href{http://dx.doi.org/10.1088/0741-3335/55/1/014010}{http://dx.doi.org/10.1088/0741-3335/55/1/014010}.
%\newpage

%\title[Influence of $\tilde{n}_e$ on microwave propagation v0.6]{Influence of plasma turbulence on microwave propagation -- v0.6}
\title[Influence of $\tilde{n}_e$ on microwave propagation]{Influence of plasma turbulence on microwave propagation}

\author{A~K\"{o}hn$^{1}$, E~Holzhauer$^2$, J~Leddy$^3$, M~B~Thomas$^3$, R~G~L~Vann$^3$}

\address{$^1$Max Planck Institute for Plasma Physics, D-85748 Garching, Germany}
\address{$^2$Institute of Interfacial Process Engineering and Plasma Technology, University of Stuttgart, D-70569 Stuttgart, Germany}
\address{$^3$York Plasma Institute, Department of Physics, University of York, York, YO10 5DD, UK}

\ead{alf.koehn@ipp.mpg.de}
\begin{abstract}
It is not fully understood how electromagnetic waves propagate through plasma \added{density} fluctuations when the size of the fluctuations is comparable with the wavelength of the incident radiation. In this paper, the perturbing effect of a turbulent plasma density layer on a traversing microwave beam is simulated with full-wave simulations. The deterioration of the microwave beam is calculated as a function of the characteristic turbulence structure size, the turbulence amplitude, the depth of the interaction zone and the size of the waist of the incident beam. The maximum scattering is observed for a structure size on the order of half the vacuum wavelength. The scattering and beam broadening was found to increase linearly with the depth of the turbulence layer and quadratically with the fluctuation strength. Consequences for experiments and 3D effects are considered.
\end{abstract}

%52.25.Os 	Emission, absorption, and scattering of electromagnetic radiation
%52.25.Xz 	Magnetized plasmas
%52.35.Hr 	Electromagnetic waves (e.g., electron-cyclotron, Whistler, Bernstein, upper hybrid, lower hybrid)
%52.35.Ra 	Plasma turbulence
%52.40.Db 	Electromagnetic (nonlaser) radiation interactions with plasma
%52.70.Gw 	Radio-frequency and microwave measurements
% 47.11.Bc for finite difference methods
%\submitto{\PPCF}
%\maketitle

\section{Introduction}\label{s.intro}
Electromagnetic waves in the microwave regime play an indispensable role in present plasma experiments. They are used for diagnostic and heating purposes in low- and high-temperature plasmas, see e.g.\ Refs.~\cite{Hartfuss.2014,Moisan.1992,Bornatici.1983,Prater.2004}. In tokamak plasmas, for example, microwaves with a power in the MW regime are injected to stabilize so-called neoclassical tearing modes (NTMs). This instability modifies the plasma current profile by the generation of small scale magnetic islands and can ultimately lead to a disruption~\cite{LaHaye.2006}. To suppress the growth of this instability, a localized current is driven by injected microwaves, in order to restore the original current profile~\cite{Zohm.2007}. 

Microwaves emitted by the plasma, on the other hand, can be used to obtain, for example, the spatial distribution of the electron temperature or to investigate its dynamical behavior~\cite{Volpe.2003,Freethy.2013}. In all these cases, the microwaves have to traverse the plasma boundary, a region in which substantial density fluctuation levels as high as 100\,\% are observed in fusion plasmas~\cite{Zweben.2007}. These fluctuations can disturb the microwaves injected into the plasma or emitted by it and thereby lead to reduced coupling efficiencies to the plasma or to ambiguous emission measurements, respectively. %The requirement of a localized absorption of the injection microwave and subsequent current drive in a narrow region for the NTM stabilization can be spoiled by such fluctuations. 
This paper seeks to allow an estimation of the perturbing effect of density fluctuations on a traversing microwave beam. 
%To this end, full-wave simulations are performed in which the average size and strength of the fluctuations are varied.

The investigations of microwaves propagating through a \replaced{turbulent}{fluctuating} plasma dates back to the beginning of ionospheric research, when microwaves emitted by strong astronomical radio sources were detected and studied on earth after they had passed through the ionosphere~\cite{Hewish.1951} or when radio waves were emitted from ground and their reflection from the ionosphere was studied~\cite{Booker.1950}. 
\added{In the 1970s and 1980s, the scattering of electromagnetic waves at a turbulent plasma density layer was used to diagnose the turbulence itself~\cite{Surko.1983,Slusher.1980}.}
\replaced{Powerful }{In the 1980s, powerful}
microwave sources started to become available for substantial plasma heating \added{in the 1980s} and first numerical studies based on geometrical optics assumptions were performed in order to estimate the deteriorating effect of a turbulent plasma edge layer on an injected microwave beam~\cite{Hansen.1988}. Recently, this topic has gained new momentum due to the importance of effective non-inductive current drive schemes in fusion relevant plasmas~\cite{Tsironis.2009,Peysson.2011,Ram.2013,Balakin.2011,Poli.2015,Sysoeva.2015}.
%due to the inevitable requirement of the aforementioned NTM-stabilization in fusion relevant plasmas by localized electron cyclotron current drive (ECCD)~\cite{Tsironis.2009,Ram.2013,Balakin.2011}. 

The work presented here investigates the perturbing effect of a turbulent plasma density layer on a traversing microwave beam by means of full-wave simulations. No restricting assumptions about the size or amplitude of the turbulent \added{density} structures therefore need to be made. The full-wave treatment is required due to the large fluctuation levels that occur in fusion edge plasmas and due to the size of the density structures which can be on the order of the wavelength of the microwave.
The properties of the turbulence are varied in a series of parameter scans in order to identify the most perturbing case.
Although full-wave simulations can be quite resource-demanding in terms of computational power and memory, their application has been increased recently to cases in magnetized plasmas where geometrical optics assumptions generally fail~\cite{Koehn.2011PoP,daSilva.2010,Wright.2010,Blanco.2013,Heuraux.2014,Heuraux.2015}. 
This has been made possible due to the continuously increasing computational power~\cite{Mack.2011} and the creation of large scientifically focused computational cluster facilities. Understanding and ensuring numerical stability of full-wave codes for cases with strong density gradients in magnetized plasmas is a topic of ongoing research~\cite{daSilva.2015}. A significant amount of that research deals with reflectometry, where an injected low-power microwave beam is \emph{reflected} by the turbulent density layer (instead of being \emph{transmitted} like in the cases mentioned in the beginning) and used to probe the turbulence layer itself~\cite{Conway.2006,Lechte.2009}.
For the scenarios considered here (\added{electron} density below \added{O-mode} cut-off and moderate density gradients), stability of the simulations is not an issue. This work is the continuation of a preceding study which dealt with the influence of single blob-like structures on a traversing microwave beam~\cite{Williams.2014,Koehn.2015}.

The paper is organized as follows: details about the simulation technique and the density fluctuations used in the simulations are described in Sec.~\ref{s.numerics}. The data analysis methods are presented in Sec.~\ref{s.analysis}. In Sec.~\ref{s.results}, the simulation results are discussed and in Sec.~\ref{s.consequences}, their consequences for two experimental cases are outlined. The summary given in Sec.~\ref{s.summary} concludes the paper.

\section{Simulation details}\label{s.numerics}
This section briefly describes the codes that have been used to simulate the plasma microwave interaction and how the turbulent density profiles were generated.

\subsection{The full-wave code IPF-FDMC}\label{s.ipf-fdmc}
IPF-FDMC is a full-wave code based on the finite-difference time-domain scheme~\cite{Taflove.2000}. It solves Maxwell's equations (\ref{e.maxwell1}), (\ref{e.maxwell2}) and the fluid equation of motion for the electrons, Eq.~(\ref{e.ipffdmc_J}), on a 2D Cartesian grid. Hence, it is based on a cold plasma description. The code has been used previously to study the O-SX mode conversion process and to investigate microwave heating in general (see Refs.~\cite{Koehn.2008,Koehn.2011PoP,Koehn.2010}). 
\begin{eqnarray}
	\frac{\partial}{\partial t}\mathbf{B} &=& -\nabla\times\mathbf{E}\label{e.maxwell1}\\
	\frac{\partial}{\partial t}\mathbf{E} &=& c^2\nabla\times\mathbf{B}-\mathbf{J}/\epsilon_0 \label{e.maxwell2}\\
	\frac{\partial}{\partial t}\mathbf{J} &=& \epsilon_0\omega_{pe}^2\mathbf{E} 
		- \omega_{ce}\mathbf{J}\times\mathbf{\hat{B}}_0 - \nu_e\mathbf{J}\label{e.ipffdmc_J}
\end{eqnarray}
with \added{$c$ the speed of light, } 
$\omega_{pe}$ the electron plasma frequency 
\added{which is connected directly with the electron density $n_e$ as $\omega_{pe}=\sqrt{n_ee^2/(\epsilon_0 m_e)}$ ($e$ is the elementary charge, $\epsilon_0$ the vacuum permittivity, $m_e$ the electron mass)}, 
\replaced{$\omega_{ce}=|e|B_0/m_e$}{$\omega_{ce}$} 
the electron cyclotron frequency and 
\replaced{$\mathbf{\hat{B}}_0$ the unit vector into the direction of}{$\mathbf{B}_0$} the background magnetic field. An electron collision frequency $\nu_e$ is included in Eq.~(\ref{e.ipffdmc_J}) to resolve the energy accumulation which occurs if an X-mode encounters the upper-hybrid resonance.
\added{Obviously, the electromagnetic waves interact only with the electron density in this model which is a reasonable simplification due to the much larger mass of the ions.}

%If, in the simulations, the microwave is incident on a cold plasma resonance (the upper-hybrid resonance), the wave energy starts to accumulate there. To maintain numerical stability, an electron collision frequency $\nu_e$ is included, as can be seen in Eq.~(\ref{e.ipffdmc_J}). For the sake of simplicity, a constant value of $\nu_e$ has been used across the whole grid, which is, however, small enough to not affect the microwave outside of the resonance region. 

\subsection{The full-wave code EMIT-3D}\label{s.emit3d}
The EMIT-3D code~\cite{Williams.2014} is also based on the finite-difference time-domain scheme. Like IPF-FDMC, it is capable of simulating microwaves propagating through a cold magnetized plasma. It solves the equations, however, in a 3D geometry. Due to the increased computational resources when going to a 3D treatment, extensive parameter scans are performed with the 2D code IPF-FDMC only. This is a valid simplification for the case of perpendicular injection (with respect to the background magnetic field) since the turbulence is essentially 2D in nature (see Sec.~\ref{s.density} for details). For a number of cases, the results from both codes are compared. Dedicated scenarios which require a 3D treatment are also presented and discussed.

\subsection{The injected microwave beam}\label{s.injected_beam}
The injected microwave is modelled as a Gaussian beam with the beam waist located in the antenna plane. Neglecting the phase terms, the electric field distribution in the antenna plane reads $E(x) = \exp\left( - x^2/w_0^2\right)$, with $x$ the coordinate across the antenna and $w_0$ the spot size of the beam, i.e.\ the radius at the beam waist. The default value for the simulations is $w_0=2\,\lambda_0$ with $\lambda_0$ the vacuum wavelength of the injected microwave. Despite the relatively small beam waist, the divergence of the beam is negligible for the default size of the computational grid, which is $10\,\lambda_0\times5\,\lambda_0$. The interaction with the density turbulence occurs thus when the wave fronts of the microwave are still parallel, like in fusion plasmas, where the beam waists are usually larger by an order of magnitude. 
This grid has been chosen in order to be able to perform the required ensemble averaging (as discussed in the following section) within a reasonable computational time. 

%If not mentioned explicitly, the spot size of the beam, i.e.\ the radius at the beam waist, is $w_0=2\,\lambda_0$ with $\lambda_0$ the vacuum wavelength of the injected microwave. 
An O-mode is injected at the bottom of the 2D grid and the background magnetic field $\mathbf{B}_0$ is directed perpendicular to it. The normalized magnetic field is set to a constant value of $Y=\omega_{ce}/\omega_0=0.5$, where $\omega_0$ is the angular frequency of the injected microwave. The boundaries of the simulation grid were set to be non-radiating.

\subsection{The turbulent plasma density profiles}\label{s.density}
\begin{figure}%[t]
    \center
    \includegraphics[width=.9\textwidth]{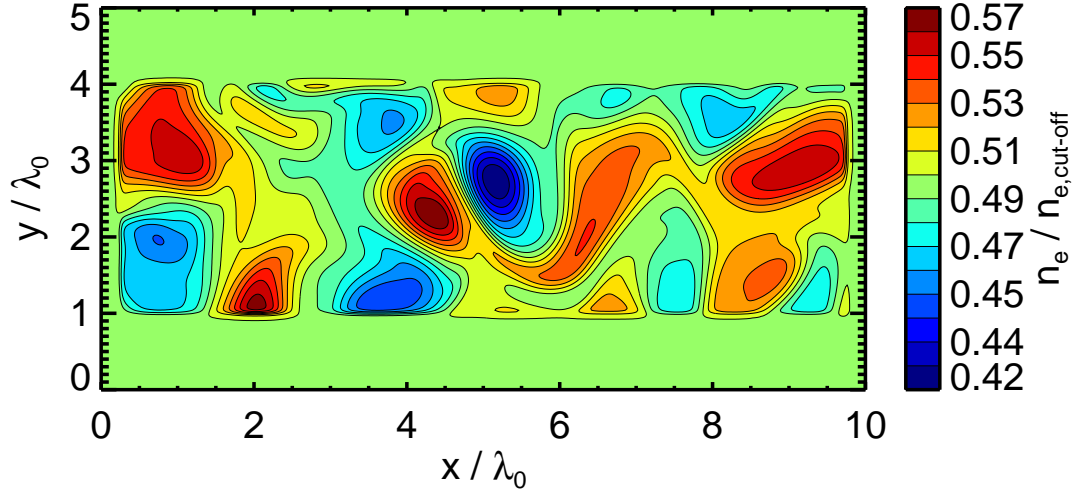}
    \caption{Contour plot of the \replaced{electron}{plasma} density normalized to the \added{O-mode} cut-off density of the injected microwave for one of the grids which is used in the full-wave simulations. The microwave is injected from the bottom and detected at the top.}
    \label{f.fullwave_grid}
\end{figure}
\added{From the turbulent plasma, only the fluctuating electron density is seen by the microwave.}
On the time scale of the microwave, the \deleted{plasma density} fluctuations appear to be frozen: the typical frequency scales \deleted{for the fluctuations} are in the kHz range~\cite{Zweben.2007,Surko.1983}, whereas the microwave frequency lies in the GHz range. The important parameter is the group velocity of the injected O-mode which can be expressed as \replaced{$v_g\approx c\sqrt{1-X}$}{$v_g\approx c_0\sqrt{1-X}$}, with \added{$c$ the speed of light and} $X=\omega_{pe}^2/\omega_0^2$. If the turbulent structures\replaced{, i.e.\ the non-homogeneities in the electron density}{ in the density profiles used here} are interpreted as drift-wave structures, their phase velocity can be estimated by the electron diamagnetic drift velocity, yielding values which are typically $\le10^4\,\mathrm{m}/\mathrm{s}$~\cite{Zweben.2007}. Hence, even if the \added{electron} density reaches values of 99.9\,\% of the \added{O-mode} cut-off density, the resulting group velocity of the wave is still a factor of 1000 larger than the velocity of the density structures and the assumption of a frozen plasma in the frame of the microwave is therefore justified.
%o-mode dispersion: w^2=w_p^2+k^2*c_0^2 --> v_g=dw/dk=k*c_0^2/w | N=c/v_ph=ck/w=sqrt(1-X) => v_g=N*c_0=c_0*sqrt(1-X)
% for the small interaction regions of only a few wavelengths size considered here. 

To properly study the influence of \replaced{electron}{plasma} density fluctuations on traversing microwaves, it is necessary to have a sufficiently large number of turbulent density profiles, since only averaging over an ensemble of profiles makes the results statistically relevant. A Hasegawa-Wakatani drift-wave turbulence model within the BOUT++ framework~\cite{Dudson.2009} is used to generate such a set of profiles which can be accessed at~\cite{Leddy.2016}.
Drift-wave turbulence is thought to be the dominant mechanism responsible for the anomalous transport observed in the tokamak plasma edge~\cite{Horton.1990,Wootton.1990}. Hasegawa and Wakatani derived a simple model to describe this turbulence in a 2D unsheared magnetic field~\cite{Wakatani.1984}, and it has been used both to simulate edge turbulence as well as to benchmark more complex models~\cite{Koniges.1992,Scott.1991,Camargo.1995}.

The geometry in the full-wave simulations is such that a homogeneous background \added{electron} density $n_0$ is chosen onto which a layer of turbulent \added{electron} density \added{fluctuations} is superimposed. Figure~\ref{f.fullwave_grid} shows an example, where the turbulence layer has a depth of $d_{\subtxt{turb}}=3\,\lambda_0$ in the $y$-direction. The spatial extension into the $x$-direction is $10\,\lambda_0$ and will not be varied. A smooth transition is employed over a few grid points between the homogeneous background plasma and the turbulence layer. 
\added{When 3D simulations are performed, the electron density structures are taken to be uniform along the direction of the background magnetic field resembling the filamentary structures observed in experiments~\cite{Zweben.2007,Surko.1983}.}
%ensures that no discontinuity appears on the density grid which could lead to numerical instability of the full-wave code. 
The emitting antenna injecting a Gaussian beam is placed at the lower boundary of the grid and a receiving antenna at the upper boundary of the grid. Both antennas extend along the whole $x$-direction.  The distance between the antennas and the boundary of the turbulence layer is set to one vacuum wavelength $\lambda_0$, i.e.\ in Fig.~\ref{f.fullwave_grid}, the emitting antenna is defined at $y=0$ and the receiving antenna at $y=5\,\lambda_0$.

\section{Data analysis}\label{s.analysis}
\begin{figure}%[t]
    \center
    \includegraphics[width=.7\textwidth]{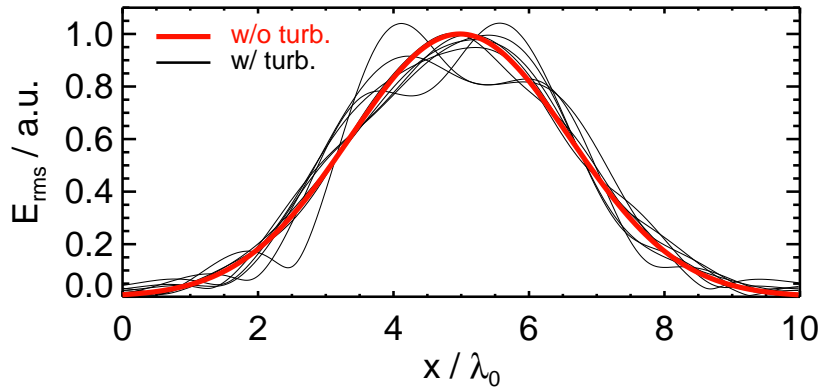}
    \caption{Signal in the detector antenna plane, a few realizations for one set of turbulence parameters (black) and the homogeneous case (red).}
    \label{f.detAnt_signal}
\end{figure}
Since the simulations are based on a time-dependent scheme, they start with the injection of the microwave beam and are stopped when a steady state solution has been achieved. Over the receiving antenna, the time-averaged wave electric field, reminiscent of an rms-value, is continuously recorded as a function of the coordinate $x$: 
%\begin{equation}
%	\tilde{E}_{\subtxt{rms}}=\sum_t\frac{|\tilde{\mathbf{E}}|} {\sqrt{T}},
%\end{equation}
\begin{equation}
	\tilde{E}_{\subtxt{rms}}=\sum_t\frac{\sqrt{ \tilde{E}_x^2 + \tilde{E}_y^2 + \tilde{E}_z^2}} {\sqrt{T}},
\end{equation}
where $t$ is the time coordinate, $T$ \replaced{the number of the wave periods passed since the start of the simulation}{the wave period} and the superscript\ \ $\tilde{ }$\ \ denotes the perturbation due to the \replaced{turbulence layer}{turbulent plasma density}. Such an antenna signal is acquired for each turbulence slice considered. Figure~\ref{f.detAnt_signal} shows this signal for a few turbulence slices together with the signal for the homogeneous, i.e.\ the unperturbed case. The effect of the turbulence on the microwave can be clearly seen. 

To quantify the scattering of the microwave beam, two different methods are used. The first method consists of calculating a scattering parameter $\alpha$ as the sum of the squared deviations of the $\tilde{E}_{\subtxt{rms}}$ signal between the turbulence case and the homogeneous case:
%This signal is then compared with the \emph{unperturbed} case, where the injected microwave is propagating along a grid with a fully homogeneous plasma with a constant density value $n_0$. To quantify the difference, a scattering parameter $\alpha$ is calculated as the sum of the squared deviations of the rms-signal at the receiving antenna:
\begin{equation}
	\alpha = \frac{\sum_x\left( \tilde{E}_{\subtxt{rms}} - E_{\subtxt{rms}} \right)^2} {\sum_x E_{\subtxt{rms}}^2}.
\label{e.alpha}
\end{equation}
This compresses the result from one turbulence slice into a single value. Ensemble averaging is then performed over many turbulence samples in order to become statistically relevant. The number of samples $N$ in one ensemble depends on the average spatial size of the \added{electron} density structures: if \replaced{they}{the structures} are very small compared to the vacuum wavelength, many of them exist in one turbulence slice resulting already in a good average. When, in contrast, the spatial dimensions of the \added{electron} density structures exceed $\lambda_0$ and only very few of them exist in one turbulence slice, a high number of samples is required. Note that according to Eq.~\ref{e.alpha}, $\alpha\ge 0$ is always true and $\alpha$ is therefore not expected to follow a normal distribution. The median and the interquartile range describing the range in which 50\% of the data is located (see e.g.\ Ref.~\cite{Moore.1993} for details) are thus chosen to characterize this scatter parameter. 

For the second analysis method, the $\tilde{E}_{\subtxt{rms}}$ signals at the receiving antenna are averaged over one ensemble of \added{electron} density profiles. A Gaussian is then fitted to the averaged signal and the beam size is determined and compared with the unperturbed beam. Thus, a measure for an effective beam broadening is obtained. The beam width is an important parameter for the earlier mentioned case of a localized current drive: if the beam is too large compared to the width of the magnetic island it is supposed to stabilize, its impact on it is reduced~\cite{Poli.2015}. The drawback of this method is, however, that side lobes due to strong scattering events are not taken into account.
%Another important quantity is described by the third analysis method, which counts the number of local maxima in the rms-signal at the receiving antenna. In Figure~\ref{f.detAnt_signal} it can be seen that the rms-signals exhibit multiple local maxima for that specific case. The occurrence of these local maxima is equivalent to a splitting of the injected microwave beam into separate beams. This splitting corresponds to an effective modulation of the injected beam and happens on the time scale it takes for the density structures to cross the microwave beam. It might have severe consequences when microwaves with a power of MWs are injected into the plasma. 

\begin{figure}%[t]
    \center
    \includegraphics[width=.7\textwidth]{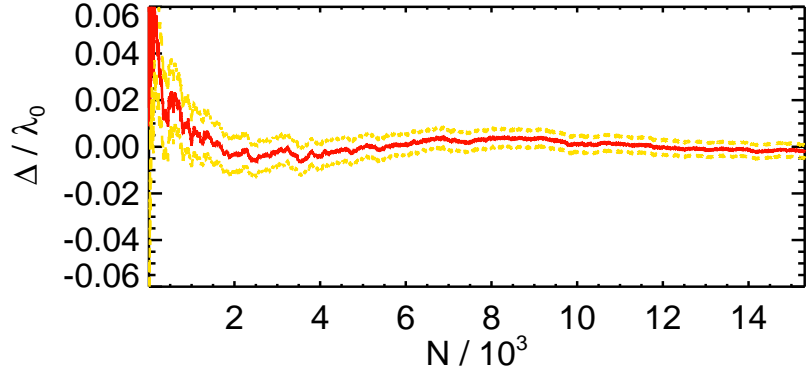}
    \caption{Average value of the spatial deviation of the maximum signal in the detector antenna plane as a function of the number of ensembles (the red central curve corresponds to the arithmetic mean and the upper and lower orange dashed curves correspond to the error given by the standard deviation of the mean).}
    \label{f.spatial_dev_max}
\end{figure}
The quality of the ensemble average is controlled by determining the average position $\tilde{x}_{\subtxt{max}}$ of the maximum in the $\tilde{E}_{\subtxt{rms}}$ signal in the detector antenna plane. This value is then compared with the position of the maximum signal $x_{\subtxt{max}}$ for the homogeneous plasma. The difference of these two values is required to be zero within the errorbars:
\begin{equation}
	\Delta=\tilde{x}_{\subtxt{max}}-x_{\subtxt{max}} \stackrel{!}{=} 0.
	\label{e.dev_pos_max}
\end{equation}

Figure~\ref{f.spatial_dev_max} shows the average of $\Delta$ for the case of \added{electron} density structures having an average size of $L_c=0.66\,(\pm0.03)\,\lambda_0$ as a function of the ensemble size. The structure size $L_c$ is defined as the distance where the spatial auto-correlation drops to a value of $0.5$. Since the magnetic field is directed perpendicular to the simulation grid, $L_c$ is also referred to as perpendicular correlation length. As can be seen in the figure, the average of $\Delta$ is deviating from zero for small values of $N$. With increasing ensemble size, $\Delta$ approaches zero and for $N=15,000$, an asymptotic value of $\Delta=0$ is reached. It has been checked that for all the cases considered in this paper,  Eq.~(\ref{e.dev_pos_max}) is fulfilled within the errorbars given by the standard deviation of the mean.

The strength of the fluctuations is calculated as the standard deviation of the \added{electron} density in the turbulence slice:
\begin{equation}
	\sigma=\sqrt{ \frac{1}{N_{x,y}} \sum_{x,y} \left( \tilde{n}_e(x,y)-n_0 \right)^2 },
	\label{e.flucLevel}
\end{equation}
where $N_{x,y}$ is the number of grid points in the turbulence slice.

\section{Results}\label{s.results}
In this section, the results of the full-wave simulations are presented. The influence of the perpendicular correlation length $L_c$ will be discussed first. Then, the effect of varying the depth of the turbulence region $d_{\subtxt{turb}}$ will be shown. The result of changing the average fluctuation strength $\sigma$ is presented thereafter, followed by an analysis of the role of the beam size $w_0$ and of the impact of varying the background density $n_0$. Finally, 3D effects occurring when injecting at a non-perpendicular angle are outlined.

\subsection{Variation of the average \replaced{size of the turbulent structures}{density structure size}}\label{s.results_rho2lambdaScan}
\begin{figure}%[t]
    \center
%	\begin{tabular}{@{}c@{}c@{}}
%	\end{tabular}
    \includegraphics[height=.3\textwidth]{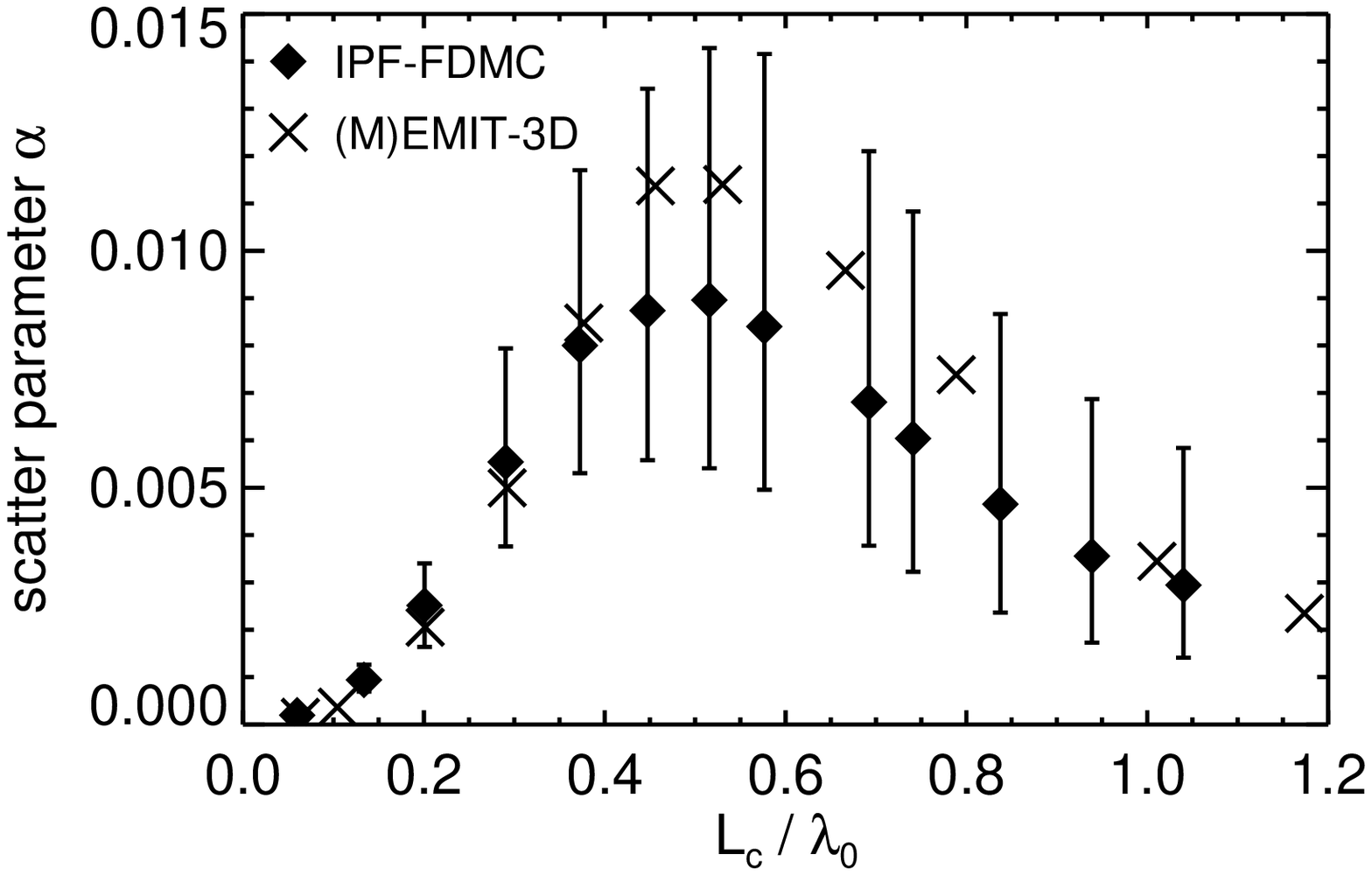}
    \includegraphics[height=.3\textwidth]{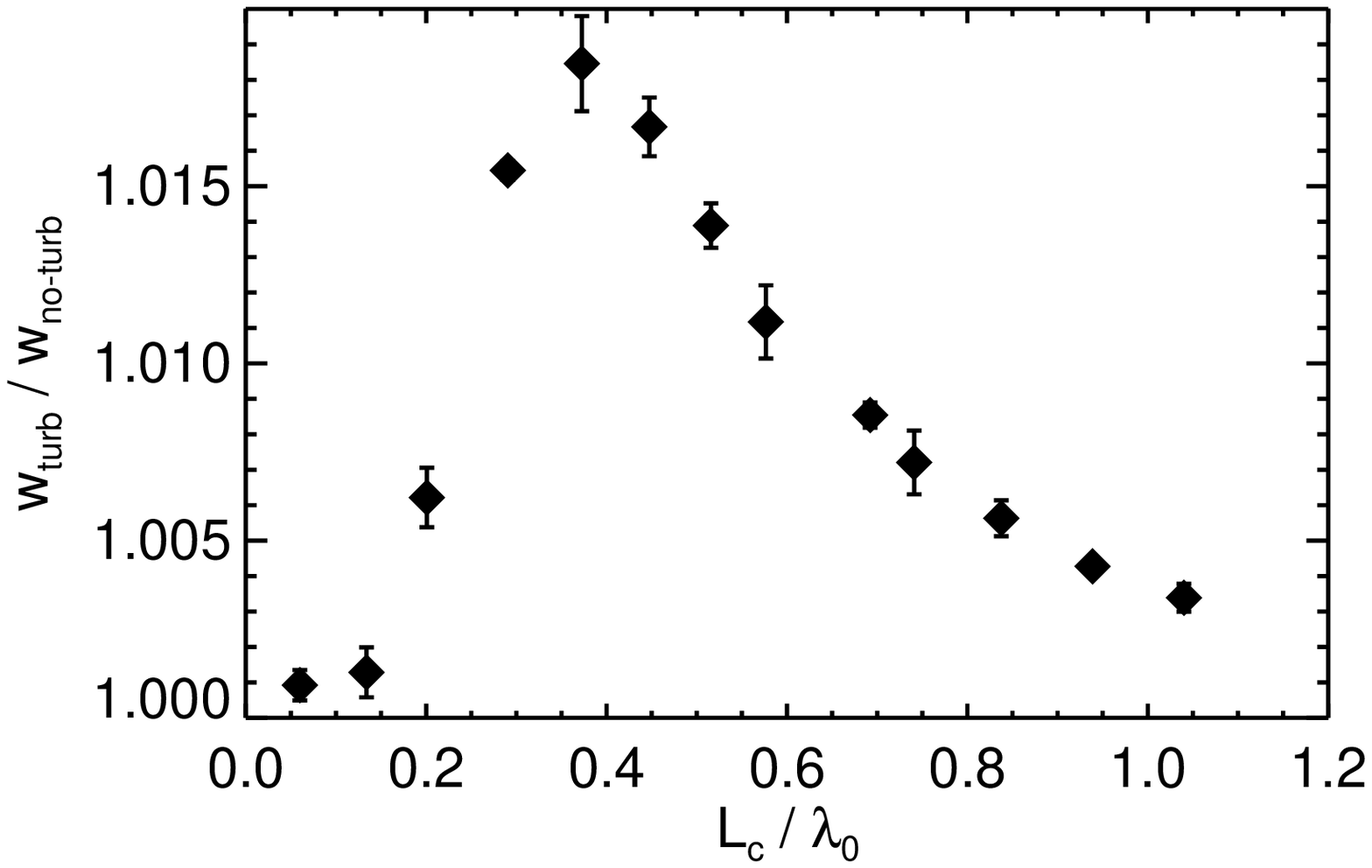}
    \caption{\emph{(Left)} Median of the scattering parameter $\alpha$ with the interquartile range as error bars and \emph{(right)} average beam broadening as a function of the average structure size $L_c$ for the case of an average fluctuation level of $\sigma\approx4\,\%$ and a size of the turbulence layer of $d_{\subtxt{turb}}=3\,\lambda_0$.}
    \label{f.cL_scan}
\end{figure}

The average size of the \added{electron} density structures in the turbulence region is varied from $L_c\approx0.06\,\lambda_0 \ldots 1.04\,\lambda_0$. 
%This is the range which requires a full-wave description of the interaction as the plasma density varies considerably over the distance of a wavelength. 
As discussed in Sec.~\ref{s.analysis}, the ensemble size $N$ increases with increasing values of $L_c$. 
%For $L_c=0.20\,(\pm 0.01)\,\lambda_0$, for example, it is $N=1200$, whereas for $L_c=1.040\,(\pm 0.003)\,\lambda_0$, it is $N=29,000$.
For $L_c\approx0.20\,\lambda_0$, for example, it is $N=1,200$, whereas for $L_c\approx1.04\,\lambda_0$, it is $N=29,000$.
The average fluctuation strength here is $\sigma=4.0\,(\pm 0.1)\,\%$ and the background density $n_0=0.5\,n_{\subtxt{e,cut-off}}$. 

As can be seen from Fig.~\ref{f.cL_scan}, the scattering (see Eq.~(\ref{e.alpha})) is found to increase with increasing structure size until a maximum of $\alpha\approx0.009$ is reached. With a further increase in structure size $L_c$, the average perturbing effect of the turbulence starts to decrease. The maximum effect on the traversing microwave beam is observed for an average structure size of $L_c=0.52\,(\pm 0.02)\,\lambda_0$. For very large \added{electron} density structures exceeding the vacuum wavelength, $L_c\gg\lambda_0$, the scattering is expected to reach an asymptotic value. For the microwave, the \replaced{turbulence}{turbulent density} layer will then appear as a homogeneously increased or decreased \added{electron} density layer corresponding to a phase plate.
Compared to the case without fluctuations, this will result in increased or decreased divergence of the beam, respectively. The error bars of $\alpha$ correspond to the interquartile range, as mentioned in Sec.~\ref{s.analysis}. Their asymmetric shape with respect to the median value reveals the skewed distribution of $\alpha$ which actually follows a log-normal distribution. They also give an indication on the spread of the data which is the strongest for the largest median of the scattering parameter. The plot shows not only data from 2D full-wave simulations performed with IPF-FDMC but also from 3D simulations performed with EMIT-3D. Due to the concomitant increase in computational time when going to 3D, the size of the ensemble of \added{electron} density profiles is reduced to $N_{\subtxt{EMIT-3D}}=128$ (independent of the value of $L_c$). Nevertheless, good agreement is found showing that this scattering process is essentially of 2D nature, as expected.
%as compared to the background density leading simply to a defocussing or focussing of the beam, respectively.  
% $\lim_{L_c\rightarrow\infty}\alpha=0$.

The scaling of the average beam broadening with $L_c$ is qualitatively similar: a maximum perturbing effect at a certain structure size is found. This size is, however, slightly smaller than in the case of the scattering parameter $\alpha$. For $L_c\gg\lambda_0$, both methods approach asymptotically the homogeneous case. The maximum beam broadening found corresponds to an increase of approximately $2\,\%$ as compared to the homogeneous case. 
It should be kept in mind that this method suffers from the fact that scattering into side lobes \deleted{due to strong turbulence events} is ignored as explained in the previous Section. 
%With increasing distance to the turbulence layer, this value is expected to increase as beam splitting by small angles, resulting in a wider beam on average, is likely to be not fully detected in the simulation geometry used (due to the small distance between turbulence layer and receiving antenna). The beam broadening found can therefore be interpreted as a minimum broadening. 
The advantage of the scattering parameter $\alpha$ is that is does not suffer from this problem as it simply sums up the squared deviations from the homogeneous case. % and by that also considering side lobes.
%otherwise hardly recognizable side lobes. 

%xxxFollowingParagraphNeedsRevisionxxx: Comparing this result with a previous study in which the influence of a single blob-like structure was investigated~\cite{Williams.2014,Koehn.2015}, it appears slightly different: the perturbing effect actually starts to decrease earlier in the turbulence case. This difference is not surprising though, as in the previous study the \emph{maximum} perturbing effect of a single blob was investigated whereas here, we investigate the \emph{average} perturbation of a fully turbulent plasma density which can be interpreted as an ensemble of blobs. Obviously, both investigations are important for an actual experiment. 

\subsection{Variation of the size of the turbulence region}\label{s.results_dturb}
\begin{figure}%[t]
    \center
    \includegraphics[width=.75\textwidth]{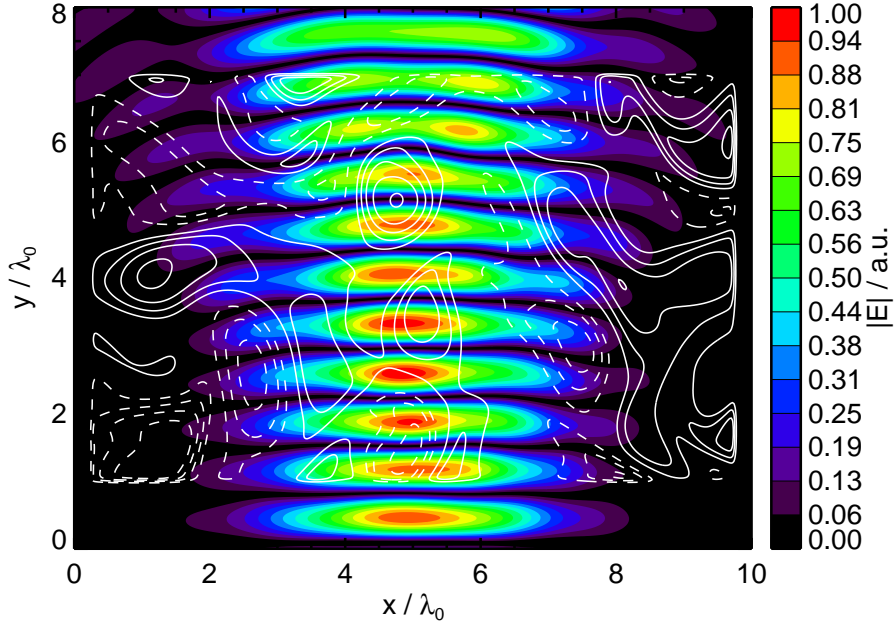}
    \caption{Contour plot of a snapshot of the absolute wave electric field together with the \added{electron} density \added{fluctuations} indicated by the white lines, where solid and dashed lines correspond to positive and negative perturbations (referring to the background \added{electron} density $n_0$), respectively. Simulation parameters are $L_c\approx0.52\,\lambda_0$, $\sigma\approx4\,\%$ and $d_{\subtxt{turb}}=6\,\lambda_0$.}
    \label{f.wavefield}
\end{figure}
The depth of the turbulent density region is varied in the range of $d_{\subtxt{turb}}=2\ldots7\,\lambda_0$. For the average structure size, the value which yielded the strongest perturbation in the previous section is used, $L_c\approx0.52\,\lambda_0$. The average fluctuation strength and the background density are the same as in Sec.~\ref{s.results_rho2lambdaScan}. 

To illustrate the perturbing effect of the \added{electron} density fluctuations on the microwave beam, a snapshot of the absolute wave electric field is shown in Fig.~\ref{f.wavefield} together with contour lines of the \added{electron} density fluctuations. The snapshot is taken after the steady state solution has been achieved. For illustration purposes, a case with a strong perturbation is chosen, $L_c\approx0.52\,\lambda_0$ and $d_{\subtxt{turb}}=6\,\lambda_0$. Looking at the top of the grid where the detector antenna plane is located, one can see how in this case the microwave beam is starting to split up into two main beams. 
%When plotting the wave electric field for the smallest average structure size considered in these simulations, $L_c\approx0.2\,\lambda_0$, hardly any perturbation is visible.

\begin{figure}%[t]
    \center
%	\begin{tabular}{@{}c@{}c@{}}
%	\end{tabular}    
    \includegraphics[height=.3\textwidth]{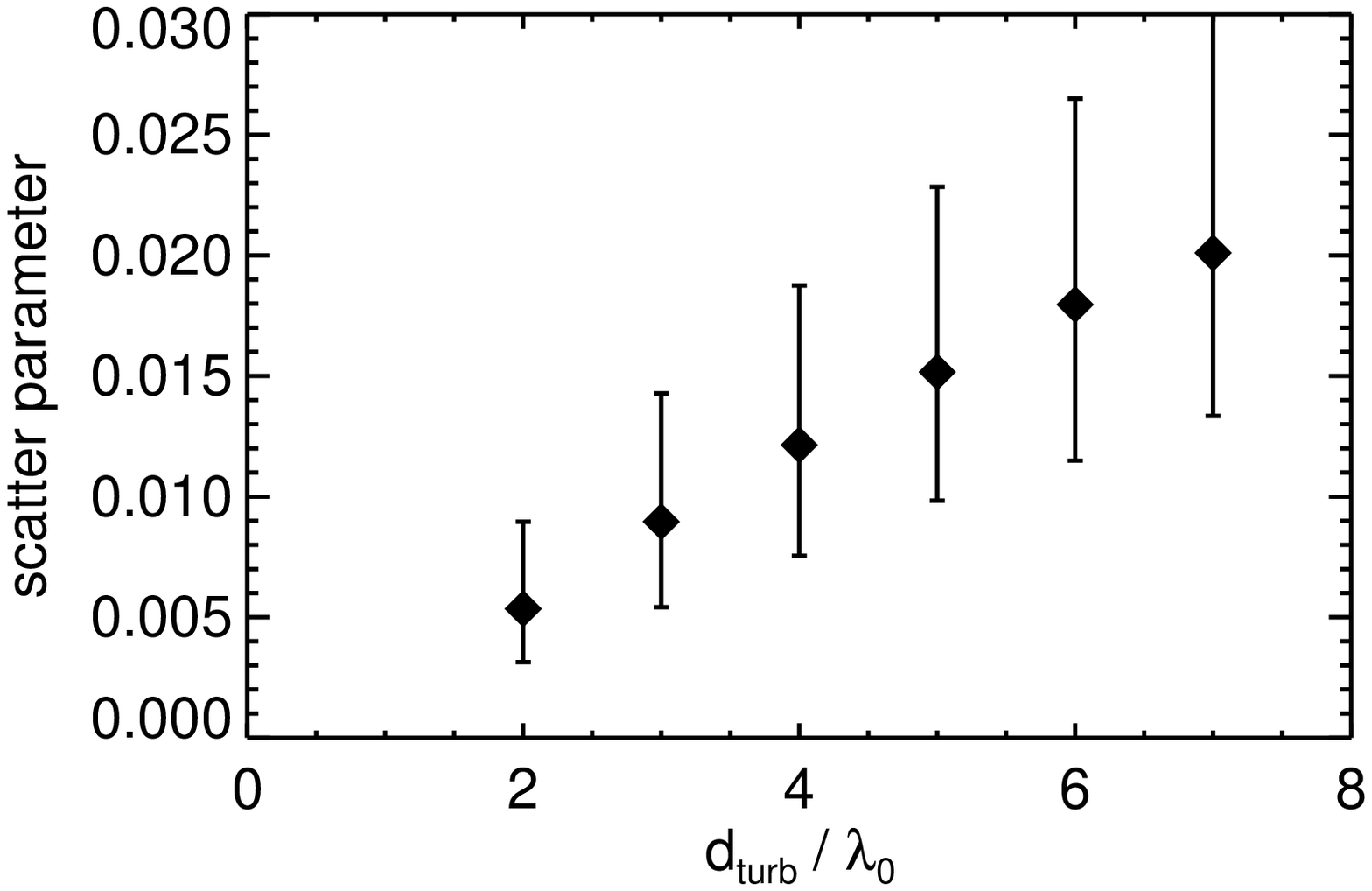}
    \includegraphics[height=.3\textwidth]{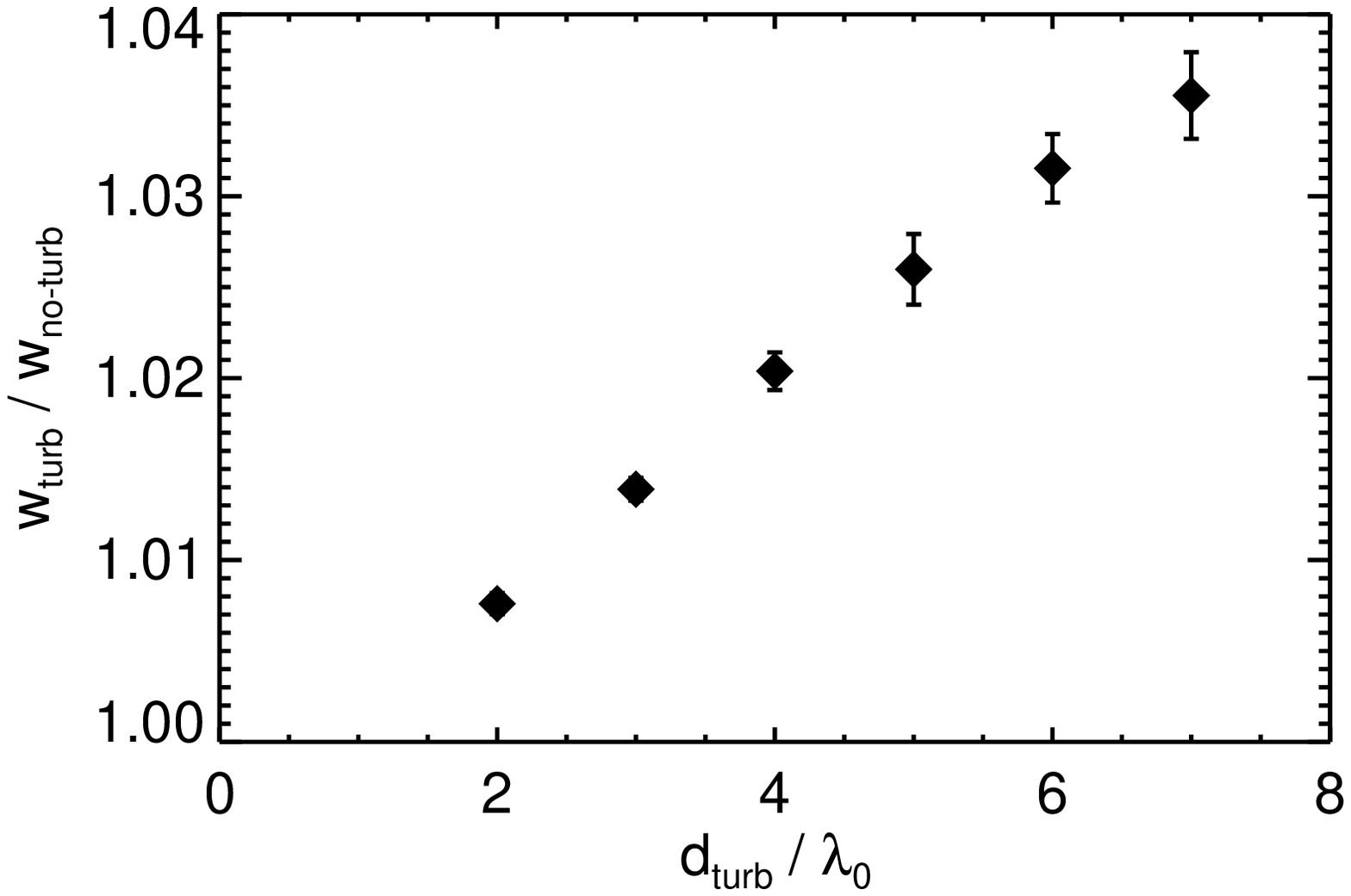}
    \caption{\emph{(Left)} Median of the scattering parameter $\alpha$ with the interquartile range as error bars and \emph{(right)} average beam broadening as a function of the thickness of the turbulence region for the case of an average fluctuation level of $\sigma\approx4\,\%$ and an average structure size of $L_c\approx0.52\,\lambda_0$.}
    \label{f.dTurb_scan}
\end{figure}

Figure~\ref{f.dTurb_scan} shows the median of the scatter parameter $\alpha$ as a function of $d_{\subtxt{turb}}$. The scattering seems to increase linearly with increasing $d_{\subtxt{turb}}$, where deviations from this behaviour can be expected for large values of $d_{\subtxt{turb}}$. Due to the increased scattering of the microwave into the left and right boundaries for interaction zones as large as the one used in Fig.~\ref{f.wavefield}, these components are not detected by the receiving antenna placed at the top of the computational domain. The average beam broadening is also shown as a function of $d_{\subtxt{turb}}$ in Fig.~\ref{f.dTurb_scan}. It exhibits a similar behaviour to the scatter parameter.

%While the first one shows an increase, the second parameter shows no clear scaling. This is again due to the too small grid or receiving antenna for this type of analysis. 

\subsection{Variation of the strength of the turbulence}\label{s.results_flucLevel}
To vary the average fluctuation amplitude, the output from the BOUT++ code, i.e.\ the mean-free \added{electron} density fluctuations, is multiplied by a constant factor before it is added to the background plasma. This procedure is chosen to ensure that the underlying turbulence is exactly the same for these full-wave simulations as it was for the other full-wave simulations. The alternative approach would be to re-run the BOUT++ simulations with different input parameters that result in larger fluctuation levels. This would have lead to a significant increase in the overall computation time. It has, however, been checked for a few cases that input parameters leading to larger fluctuation amplitudes do not change the other average turbulence parameters, namely the structure size. It is therefore justified to scale the fluctuation amplitude as it is done here. A variation of $\sigma=2\ldots12\,\%$ is realized. 

The standard parameters of $n_0=0.5\,n_{\subtxt{e,cut-off}}$ and $d_{\subtxt{turb}}=3\,\lambda_0$ are used. For the average structure size $L_c$, the value which yielded the strongest perturbation (see Sec.~\ref{s.results_rho2lambdaScan}) is again chosen. 

%\begin{figure}%[t]
%    \center
%    \includegraphics[width=.5\textwidth]{../figs/york_29.eps}
%    \caption{Scattering as a function of the average fluctuation level for the case of an average structure size of $L_c\approx0.52\,\lambda_0$ and $d_{\subtxt{turb}}=3\,\lambda_0$. The dashed line represents a power law fitted to the data.}
%    \label{f.fluclevel_scan}
%\end{figure}
\begin{figure}%[t]
    \center
%	\begin{tabular}{@{}c@{}c@{}}
%	\end{tabular}    
    \includegraphics[height=.3\textwidth]{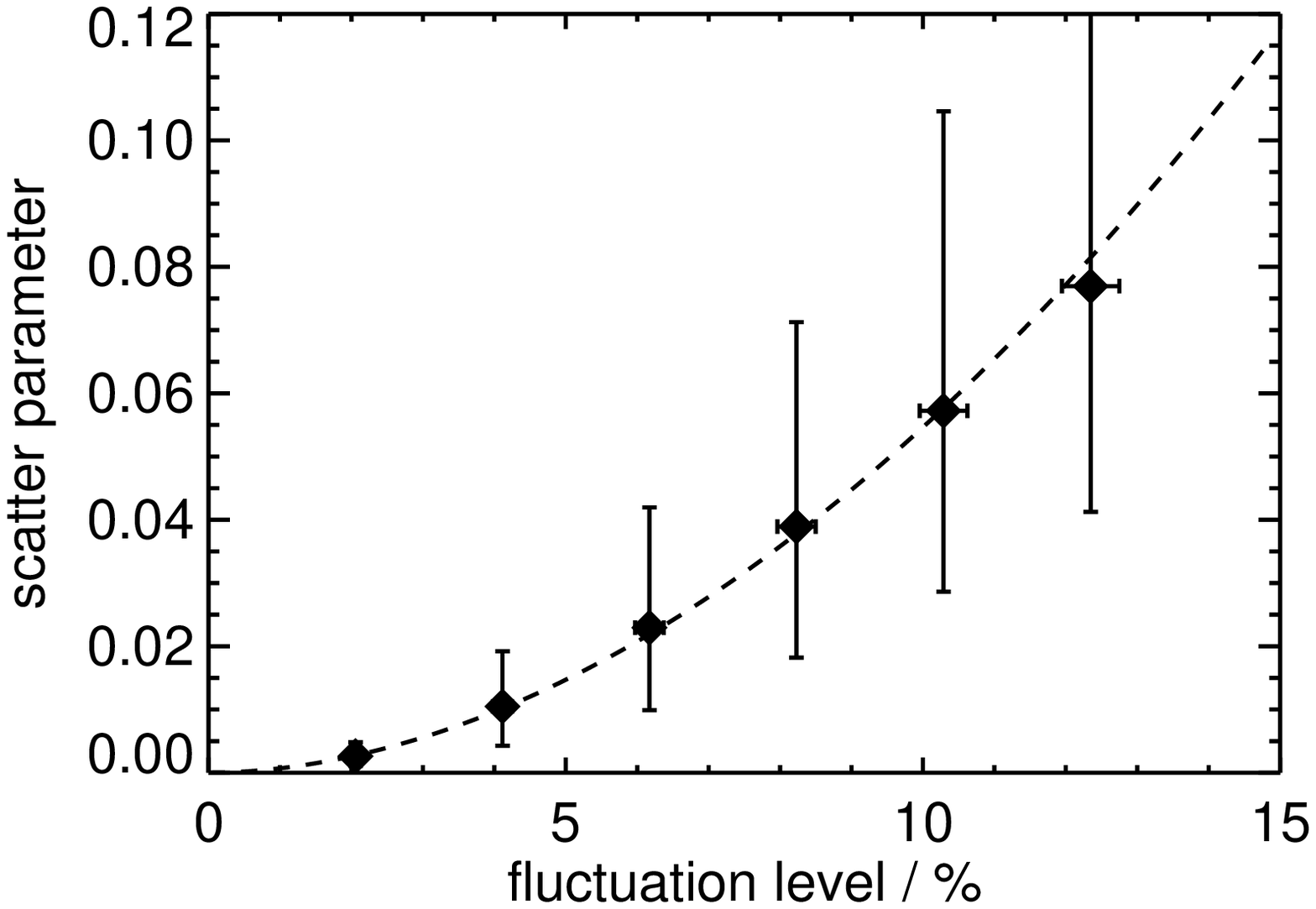}
    \includegraphics[height=.3\textwidth]{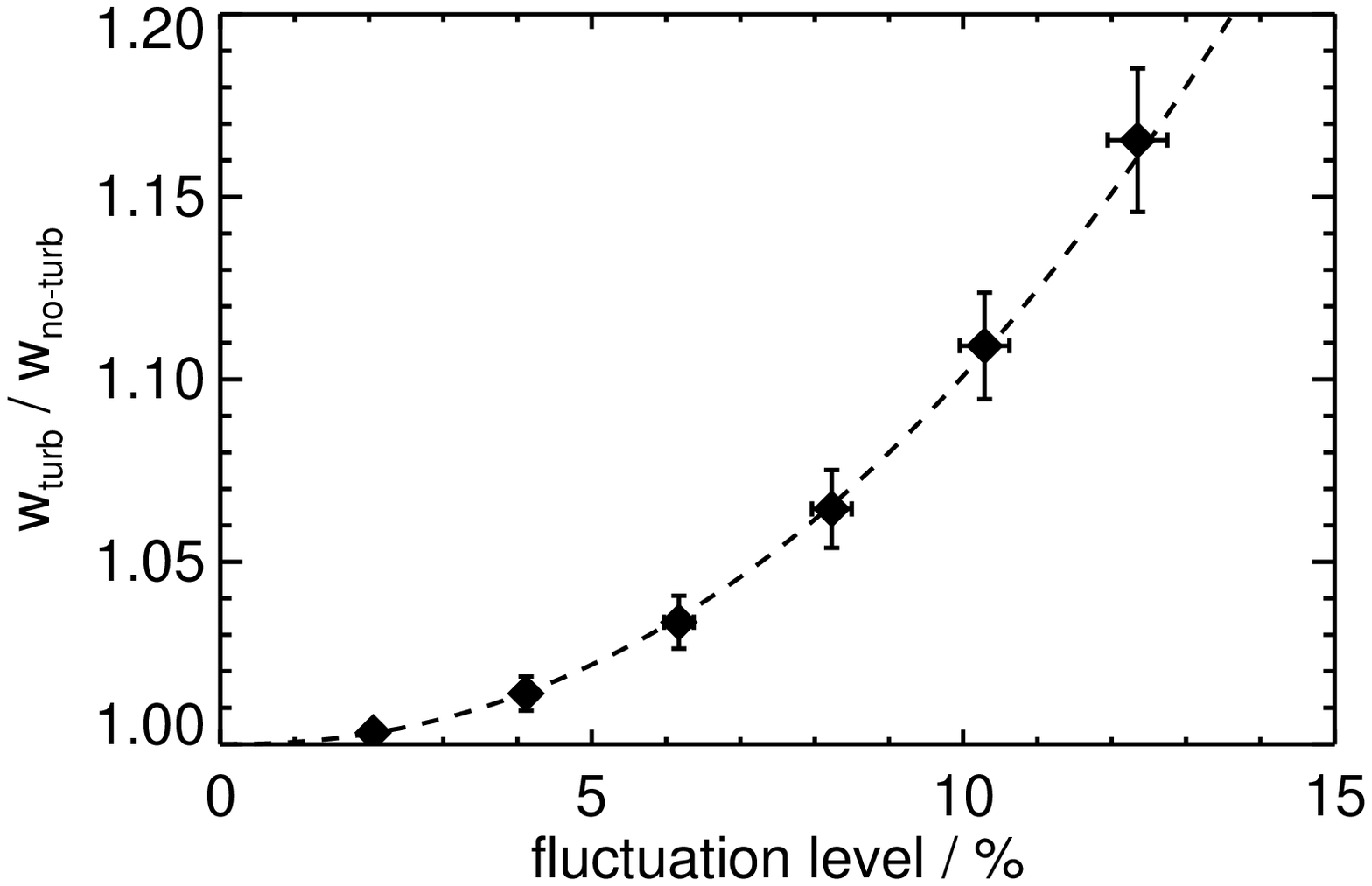}
    \caption{\emph{(Left)} Median of the scattering parameter $\alpha$ with the interquartile range as error bars and \emph{(right)} average beam broadening as a function of the normalized fluctuation amplitude for the case of an average structure size of $L_c\approx0.52\,\lambda_0$ and $d_{\subtxt{turb}}=3\,\lambda_0$. The dashed line represents a power law fitted to the data.}
    \label{f.fluclevel_scan}
\end{figure}

Figure~\ref{f.fluclevel_scan} shows the scattering parameter $\alpha$ as a function of the fluctuation level $\sigma$. The increase can be described by a power law of the form $\alpha=a\,\sigma^b$, with $a\approx1\cdot10^{-3}$ and $b\approx1.9$, i.e.\ approximately a quadratic increase. When the turbulence layer is assumed to be of small depth and treated as a phase variation of a traversing Gaussian beam, it can be shown that the intensity variation due to the phase perturbation scales with the square of the \added{electron} density variation~\cite{Evans.1982}. Hence, the parameters considered for the full-wave simulations in this section, resemble this case.

The average beam broadening exhibits the same behaviour as the scattering parameter $\alpha$, see Fig.~\ref{f.fluclevel_scan}. Its dependence on the fluctuation level can again be described by a power law with rather similar coefficients of $a\approx1.1\cdot10^{-3}$ and $b\approx2.1$.

%Important for the experimentalist is the fact that the errorbars are also increasing with increasing fluctuation strength since this means that cases with stronger deviations from the average value are to be expected. 

\subsection{Variation of the size of the beam waist}\label{s.results_w0variation}
In the description of the simulation geometry (see Sec.~\ref{s.ipf-fdmc}) it was argued that despite the relatively small value of $w_0$, the simulations still resemble fusion-relevant cases with the beam waist larger by one order of magnitude. This is due to the fact that the beam diverges only marginally in the small simulation domain. In this section, the beam waist is varied from $w_0=1\ldots6\,\lambda_0$ to investigate its influence. To contain these wide beams, the size of the computational domain is increased in the $x$ - direction to $x=30\,\lambda_0$. The other parameters are kept at their default values of $L_c\approx0.52\,\lambda_0$, $\sigma\approx4\,\%$, and $d_{\subtxt{turb}}=3\,\lambda_0$.

\begin{figure}%[t]
    \center
    \includegraphics[height=.3\textwidth]{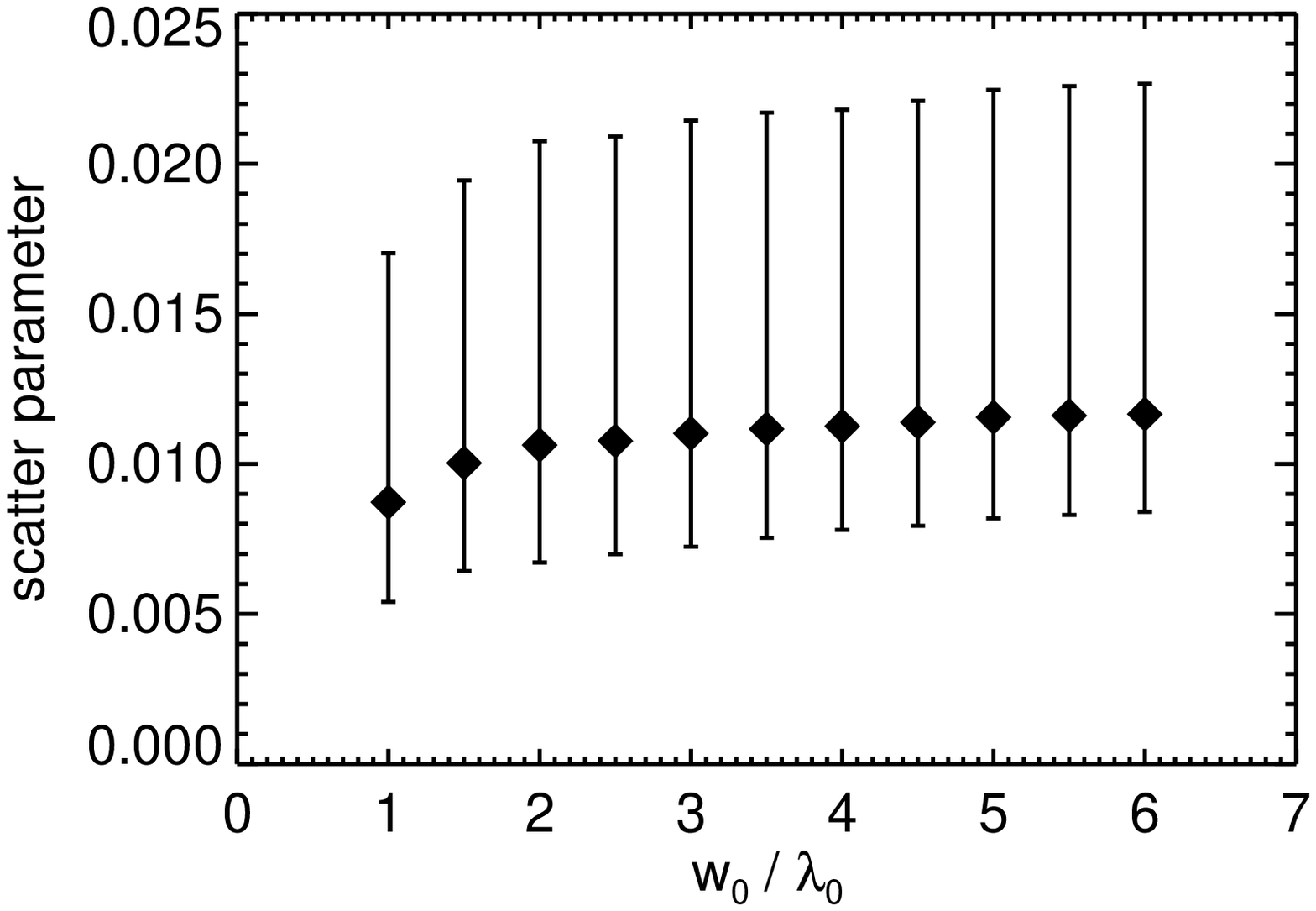}
    \includegraphics[height=.3\textwidth]{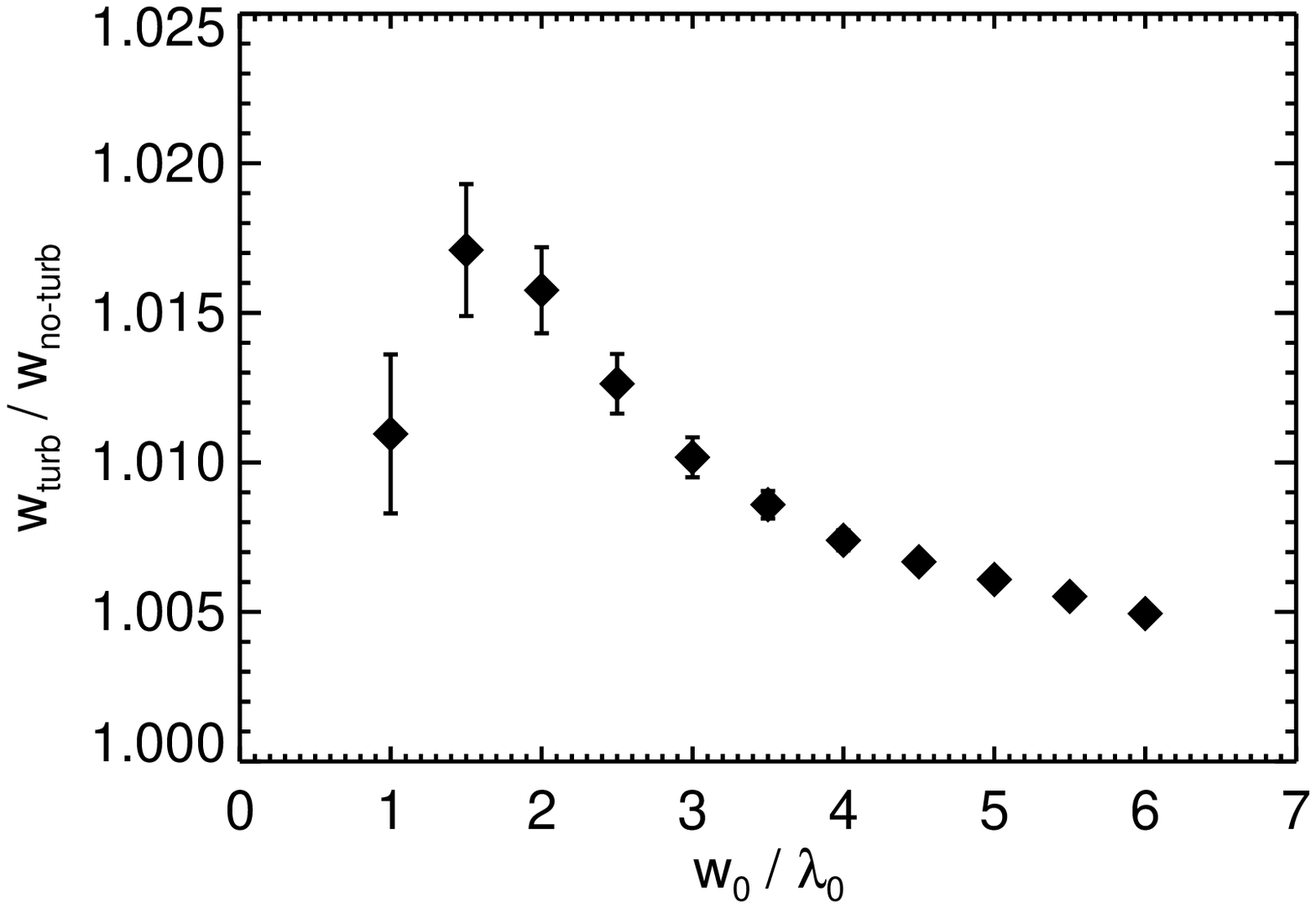}
    \caption{\emph{(Left)} Median of the scattering parameter $\alpha$ with the interquartile range as error bars and \emph{(right)} average beam broadening as a function of the size of the beam waist for the case of an average fluctuation level of $\sigma\approx4\,\%$, an average structure size of $L_c\approx0.52\,\lambda_0$, and a depth of the turbulence layer of $d_{\subtxt{turb}}=3\,\lambda_0$.}
    \label{f.w0_scan}
\end{figure}

With increasing size of the beam waist, the scattering parameter $\alpha$ approaches an asymptotic value, as can be seen from Fig.~\ref{f.w0_scan}. The asymptotic value is approximately reached for $w_0=2\,\lambda_0$, which corresponds to the default value used in the parameter scans. The average beam broadening exhibits a slightly different behaviour. It also approaches an asymptotic value but coming from higher values this time. This is due to the fact that the average beam broadening is a normalized quantity (normalized to the unperturbed beam). Thus, the wider the beam, the smaller the \emph{relative} beam broadening. The decrease of the beam broadening for the smallest beam considered can be explained by the geometry used: in the detector antenna plane, the smallest beam has a wider electric field distribution as the beam with the next beam width considered. This is due to the strong divergence for the smallest beam considered.

%It has a maximum at around $w_0=1.5\,\lambda_0$. Towards smaller beams, the average broadening seems to decrease which is due to the fact that such small beams diverge significantly across the computational domain. At the detection antenna plane a wider beam is therefore detected as for a beam with a slightly larger size of the beam waist (which only diverges marginally across the computational domain). Since the scattering is not increased, the average beam broadening seems therefore to be reduced. The more relevant case, however, is the behaviour towards larger beam waists: the average beam broadening decreases slightly towards an asymptotic value. This can be understood from the concomitant behaviour of the scattering parameter $\alpha$. Since it has already reached an asymptotic value, with increasing size of the beam waist, the average beam broadening is therefore slowly decreasing (note that the average beam broadening is given here as a normalized quantity).

\subsection{Variation of the \added{electron} background density}\label{s.results_n0variation}
In all the simulations shown so far, the background \added{electron} density was kept fixed at a value of $n_0=0.5\,n_{\subtxt{e,cut-off}}$. Now it is decreased by a factor of 5 to a value of $n_0=0.1\,n_{\subtxt{e,cut-off}}$. In this configuration, simulations were performed with $d_{\subtxt{turb}}=3\,\lambda_0$ and the average \added{electron} density structure size which yielded the strongest perturbation. An average scattering parameter of $\alpha=2.11\,(\pm 0.06)\cdot 10^{-4}$ is obtained. Note that this is a factor of 50 below the corresponding value for the higher background \added{electron} density. The average beam broadening is with a value of $w_{\subtxt{turb}}/w_{\subtxt{no-turb}}=1.0002\,(\pm3\cdot10^{-5})$ 
smaller by a similar factor and thus negligible.

\subsection{3D simulations for a beam propagating non-perpendicular to the density perturbations}\label{s.results_3D}
In the simulations presented so far, the microwave beam was injected perpendicular onto the \added{electron} density structures (and thus onto the magnetic field). In the scenario considered in this section, the 3D code EMIT-3D is used to simulate the case of an angle of $\pi/4$ between the microwave beam and the magnetic field. The goal of this section is not to present a comprehensive parameter scan including a full angular scan. Such scans would increase the computational time to an unacceptable value. The idea is rather to illustrate effects occurring for an oblique injection onto the \added{electron} density structures. 

\begin{figure}%[t]
    \center
    \includegraphics[width=.7\textwidth]{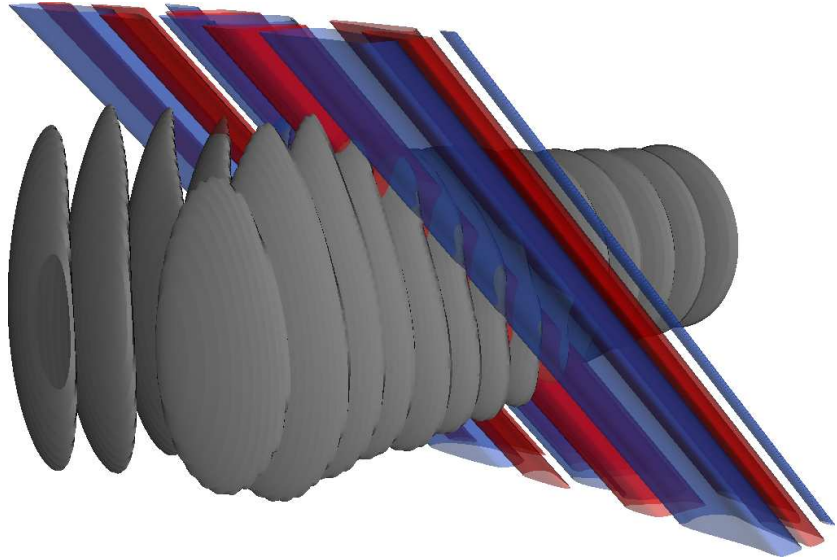}
    \caption{Snapshot of the positive wave electric field together with contours of the \added{electron} density \replaced{fluctuations}{turbulence} (red and blue corresponding to positive and negative perturbations respectively) for an angle of 45 degrees between the injected microwave beam and the \added{electron} density filaments.}
    \label{f.3Drun}
\end{figure}

Figure~\ref{f.3Drun} shows a snapshot of the wave electric field of a microwave beam in the geometry described. The size of the beam waist is $w_0=2\,\lambda_0$ in both directions perpendicular to the propagation. The \added{electron} density fluctuations are assumed not to vary in the direction parallel to the background magnetic field, thus their filamentary appearance. The scattering of the microwave beam is clearly 3D in nature. Running the simulation for a few turbulence slices, indications of a reduced scattering are found as compared to the perpendicular injection which require, however, further verification using a sufficiently good ensemble average.

\section{Consequences for the experimentalist}\label{s.consequences}
In this section, the consequences of the simulation results are briefly discussed for two different experimental cases. These are the diagnostics of microwave radiation emitted by the plasma and the localized absorption of high-power microwaves in the confinement region of the plasma.

\subsection{Microwave emission experiments}
The case of the Synthetic Aperture Microwave Imaging (SAMI) diagnostics installed at the MAST spherical tokamak is considered here~\cite{Shevchenko.2012}. This diagnostics detects electron Bernstein wave emission in a frequency range of $10\ldots36.5\,\mathrm{GHz}$, corresponding to a vacuum wavelength of $\lambda_0\approx1\ldots3\,\mathrm{cm}$. The electrostatic electron Bernstein waves are generated by electron temperature fluctuations in the electron cyclotron frequency range. They are mode-converted into electromagnetic waves in the vicinity of the plasma frequency layer and the upper-hybrid resonance layer~\cite{Laqua.2007} and can then leave the plasma. The SAMI diagnostic aims to use the recorded signal to estimate the pitch angle and thus the edge current density profile.

The case of a background density of $n_0=0.5\,n_{\subtxt{e,cut-off}}$ discussed in Sec.~\ref{s.results_rho2lambdaScan} can be applied here. The average size of the density structures at the plasma edge is $L_c=5\ldots10\,\mathrm{cm}$~\cite{Dudson.2008,Ayed.2009} and an average amplitude of $10\ldots20\,\%$ can be assumed. Comparing the structure size with the wavelength range of the diagnostics, the scattering should be noticeable in the lower frequency range (corresponding to the larger wavelength of the microwave). At the upper end of the frequency sensitivity of SAMI (corresponding to the smaller wavelength of the microwave), it should be strongly reduced. 

The large frequency range of SAMI provides an interesting scenario to experimentally investigate the scattering of microwaves as a function of the wavelength of the microwave (equivalent to varying the average density structure size). The results could then be compared with appropriate simulations.

%With this large frequency range, this diagnostics actually provides a configuration which is interesting for investigating the scattering effect experimentally and then compare with appropriate simulations.

\subsection{Microwave injection and absorption experiments}
%170GHz: cut-off=3.6e20 | 140GHz: cut-off=2.43e20
%fact of larger w_0 is not discussed
Injected microwaves in the electron cyclotron frequency range can be used to drive toroidal net currents~\cite{Erckmann.1994}. This effect can suppress the growth of magnetic islands in order to stabilize NTMs, as explained in Sec.~\ref{s.intro}. The example of the ASDEX Upgrade tokamak is considered here~\cite{Zohm.2007} with a heating frequency of $140\,\mathrm{GHz}$, corresponding to $\lambda_0\approx2\,\mathrm{mm}$ and a cut-off density of $n_{\subtxt{e,cut-off}}\approx 2.4\cdot10^{20}\,\mathrm{m}^{-3}$. In ASDEX Upgrade, the size of the blobs appearing in the scrape-off layer is on the average $L_c\approx7\,\mathrm{mm}$~\cite{Fuchert.2014}, the density there usually $n_e<3\cdot10^{19}\,\mathrm{m}^{-3} \approx 0.1\cdot n_{\subtxt{e,cut-off}}$~\cite{Stober.2000} and the fluctuation strength $\sigma\approx15\,\%$~\cite{Hennequin.2015}. 

Despite this rather large fluctuation amplitude, the effect on the injected microwave is expected to be small since the background density is far below the corresponding cut-off density. This has been discussed in Sec.~\ref{s.results_n0variation}. The distance from the antenna to the intended place of absorption is, however, very large with a value on the order of $100\,\lambda_0$. Very small scattering angles are therefore expected to be of larger significance than in the simulation geometry discussed in this paper. Appropriate modelling of such scenarios will be performed as the next step and compared with quasi-analytical approaches.

\section{Summary}\label{s.summary}
Full-wave simulations of a microwave beam injected in O-mode polarization perpendicular onto a \replaced{layer of turbulent electron density fluctuations}{turbulent plasma density layer} have been performed. In a series of parameter scans the properties of the \added{electron} density fluctuations and of the microwave beam have been varied. The strongest deterioration of the microwave beam was found for a perpendicular correlation length of the \added{electron} density structures of $L_c\approx\lambda_0/2$. The scattering of the microwave and the average beam broadening was found to increase linearly with the depth of the turbulence layer and quadratically with the fluctuation strength. The latter behaviour corresponds to the theory of scattering at a thin phase grid. An asymptotic behaviour was observed for increasing size of the beam waist. 

Very good agreement between 2D and 3D simulations was demonstrated for one  set of parameters. The effect of an obliquely injected microwave beam was illustrated for one example with a 3D simulation. Indications for an overall reduced scattering were found if the microwave beam propagates no longer perpendicular to the \added{electron} density perturbations.

Discussing the simulation results in an experimental context, it was shown that scenarios exist where the effect of density fluctuations on a propagating microwave should be taken into account. The simulations will be applied to ITER-like scenarios as a next step.

\section{Acknowledgements}
Valuable discussions with Drs.\ Omar Maj and Garrard Conway are gratefully acknowledged by one of the authors (A.~K.). Part of the simulations were performed on the HELIOS supercomputer system at Computational Simulation Centre of International Fusion Energy Research Centre (IFERC-CSC), Aomori, Japan, under the Broader Approach collaboration between Euratom and Japan, implemented by Fusion for Energy and JAEA. One of the authors (M.B.~T.) was funded by the EPSRC Centre for Doctoral Training in Science and Technology of Fusion Energy grant EP/L01663X.


\section*{References}
\begin{thebibliography}{10}
	\bibitem{Hartfuss.2014} Hartfuss H-J, and Geist T {\it Fusion Plasma Diagnostics with mm-Waves} (Wiley-VCH, Weinheim, 2014) 
	\bibitem{Moisan.1992} Moisan M, and Pelletier J {\it Microwave Excited Plasmas} (Elsevier Science Publishers B.V., Amsterdam-London-New York-Tokio, 1992) 
    \bibitem{Bornatici.1983} Bornatici M \etal 1983 {\it \NF} \href{http://dx.doi.org/10.1088/0029-5515/23/9/005}{{\bf 23} 1153}
    \bibitem{Prater.2004} Prater R 2004 {\it \PoP} \href{http://dx.doi.org/10.1063/1.1690762}{{\bf 11} 2349}
    \bibitem{LaHaye.2006} La Haye R J 2006 {\it \PoP} \href{http://dx.doi.org/10.1063/1.2180747}{{\bf 13} 055501}
    \bibitem{Zohm.2007} Zohm H \etal 2007 {\it \NF} \href{http://dx.doi.org/10.1088/0029-5515/47/3/010}{{\bf 47} 228}
    \bibitem{Volpe.2003} Volpe F \etal 2003 {\it \RSI} \href{http://dx.doi.org/10.1063/1.1530379}{{\bf 74} 1409}
    \bibitem{Freethy.2013} Freethy S \etal 2013 {\it \PPCF} \href{http://dx.doi.org/10.1088/0741-3335/55/12/124010}{{\bf 55} 124010}
    \bibitem{Zweben.2007} Zweben S J \etal 2007 {\it \PPCF} \href{http://dx.doi.org/10.1088/0741-3335/49/7/S01}{{\bf 49} S1}
    \bibitem{Hewish.1951} Hewish A 1951 {\it Proc. of the Royal Soc. of London A} \href{http://dx.doi.org/10.1098/rspa.1951.0189}{{\bf 209} 81}
    \bibitem{Booker.1950} Booker H G \etal 1950 {\it Phil. Trans. A} \href{http://dx.doi.org/10.1098/rsta.1950.0011}{{\bf 242} 579}    
	\bibitem{Surko.1983} Surko C M, and Slusher R E 1983 {\it Science} \href{http://dx.doi.org/10.1126/science.221.4613.817}{{\bf 221} 817}
	\bibitem{Slusher.1980} Slusher R E, and Surko C M 1980 {\it Phys. Fluids} \href{http://dx.doi.org/10.1063/1.863016}{{\bf 23} 472}
	\bibitem{Hansen.1988} Hansen F R \etal 1988 {\it \NF} \href{http://dx.doi.org/10.1088/0029-5515/28/5/002}{{\bf 28} 769}
	\bibitem{Tsironis.2009} Tsironis C \etal 2009 {\it \PoP} \href{http://dx.doi.org/10.1063/1.3264105}{{\bf 16} 112510}
    \bibitem{Peysson.2011} Peysson I \etal 2011 {\it \PPCF} \href{http://dx.doi.org/10.1088/0741-3335/53/12/124028}{{\bf 53} 124028}
	\bibitem{Ram.2013} Ram A \etal 2013 {\it \PoP} \href{http://dx.doi.org/10.1063/1.4803898}{{\bf 20} 056110}
	\bibitem{Balakin.2011} Balakin A A \etal 2011 {\it IEEE Trans. Plasma Science} \href{http://dx.doi.org/10.1109/TPS.2011.2158666}{{\bf 39} 3012}
	\bibitem{Poli.2015} Poli E \etal 2015 {\it \NF} \href{http://dx.doi.org/10.1088/0029-5515/55/1/013023}{{\bf 55} 013023}
	\bibitem{Sysoeva.2015} Sysoeva E V \etal 2015 {\it \NF} \href{http://dx.doi.org/10.1088/0029-5515/55/3/033016}{{\bf 55} 033016}
    \bibitem{Williams.2014} Williams T R N \etal 2014 {\it \PPCF} \href{http://dx.doi.org/10.1088/0741-3335/56/7/075010}{{\bf 56} 075010}
    \bibitem{Heuraux.2014} Heuraux S \etal 2014 {\it C. R. Physique} \href{http://dx.doi.org/10.1016/j.crhy.2014.04.004}{{\bf 15} 421}
    \bibitem{Heuraux.2015} Heuraux S \etal 2015 {\it \JPP} \href{http://dx.doi.org/10.1017/S0022377815000951}{{\bf 81} 435810503}    
    \bibitem{Koehn.2011PoP} K{\"o}hn A \etal 2011 {\it \PoP} \href{http://dx.doi.org/10.1063/1.3609828}{{\bf 18} 082501}	    
    \bibitem{daSilva.2010} {da Silva} F \etal 2010 {\it IEEE Trans. Plasma Science} \href{http://dx.doi.org/10.1109/TPS.2010.2056703}{{\bf 38} 2144}
    \bibitem{Wright.2010} Wright J C \etal 2010 {\it IEEE Trans. Plasma Science} \href{http://dx.doi.org/10.1109/TPS.2010.2055167}{{\bf 38} 2136}
    \bibitem{Blanco.2013} Blanco E \etal 2013 {\it \PPCF} \href{http://dx.doi.org/10.1088/0741-3335/55/12/125006}{{\bf 55} 125006}
    \bibitem{Mack.2011} Mack C A 2010 {\it IEEE Trans. Semiconductor Manufacturing} \href{http://dx.doi.org/10.1109/TSM.2010.2096437}{{\bf 24} 202}    
    \bibitem{daSilva.2015} {da Silva} F \etal 2015 {\it J. Comp. Phys.} \href{http://dx.doi.org/10.1016/j.jcp.2015.03.069}{{\bf 295} 24}	
    \bibitem{Lechte.2009} Lechte C 2009 {\it IEEE Trans. Plasma Science} \href{http://dx.doi.org/10.1109/TPS.2009.2019651}{{\bf 37} 1099}
    \bibitem{Conway.2006} Conway G D 2006 {\it \NF} \href{http://dx.doi.org/10.1088/0029-5515/46/9/S01}{{\bf 46} S665} 
    \bibitem{Koehn.2015} K{\"o}hn A \etal 2015 {\it EPJ Web of Conferences} \href{http://dx.doi.org/10.1051/epjconf/20158701003}{{\bf 87} 01003}        
	\bibitem{Taflove.2000} Taflove A, and Hagness S C {\it Computational Electrodynamics: the Finite-Difference Time-Domain Method} (Artech House Publishers, Boston, 2000) 
    \bibitem{Koehn.2008} K{\"o}hn A \etal 2008 {\it \PPCF} \href{http://dx.doi.org/10.1088/0741-3335/50/8/085018}{{\bf 50} 085018}
    \bibitem{Koehn.2010} K{\"o}hn A \etal 2010 {\it \PPCF} \href{http://dx.doi.org/10.1088/0741-3335/52/3/035003}{{\bf 52} 035003}	    
    \bibitem{Dudson.2009} Dudson B \etal 2009 {\it Comp. Phys. Comm.} \href{http://dx.doi.org/10.1016/j.cpc.2009.03.008}{{\bf 180} 1467}
	\bibitem{Leddy.2016} Leddy J \etal {\it Plasma density turbulence obtained from a Hasegawa-Wakatani drift-wave turbulence model within the BOUT++ framework} (2016) \href{http://dx.doi.org/10.5281/zenodo.47206}{doi:10.5281/zenodo.47206} 
    \bibitem{Horton.1990} Horton W 1990 {\it Physics Reports} \href{http://dx.doi.org/10.1016/0370-1573(90)90148-U}{{\bf 192} 1}    
    \bibitem{Wootton.1990} Wootton A J \etal 1990 {\it Phys. Fluids B} \href{http://dx.doi.org/10.1063/1.859358}{{\bf 2} 2879}	
    \bibitem{Wakatani.1984} Wakatani M \etal 1984 {\it Phys. Fluids} \href{http://dx.doi.org/10.1063/1.864660}{{\bf 27} 611}
    \bibitem{Koniges.1992} Koniges A E \etal 1992 {\it Phys. Fluids B} \href{http://dx.doi.org/10.1063/1.860151}{{\bf 4} 2785}
    \bibitem{Scott.1991} Scott B D \etal 1991 {\it Phys. Fluids B} \href{http://dx.doi.org/10.1063/1.859956}{{\bf 3} 51}
    \bibitem{Camargo.1995} Camargo S J \etal 1995 {\it \PoP} \href{http://dx.doi.org/10.1063/1.871116}{{\bf 2} 48}
	\bibitem{Moore.1993} Moore D S, and McCabe G P {\it Introduction to the Practice of Statistics} (W. H. Freeman and Company, New York, 1993) 
   	\bibitem{Evans.1982} Evans D E \etal 1982 {\it Plasma Physics} \href{http://dx.doi.org/10.1088/0032-1028/24/7/009}{{\bf 24} 819}	
    \bibitem{Shevchenko.2012} Shevchenko V F \etal 2012 {\it J. Instrum.} \href{http://dx.doi.org/10.1088/1748-0221/7/10/P10016}{{\bf 7} 1467}
    \bibitem{Laqua.2007} Laqua H P 2007 {\it \PPCF} \href{http://dx.doi.org/10.1088/0741-3335/49/4/R01}{{\bf 49} R1}
    \bibitem{Erckmann.1994} Erckmann V, and Gasparino U 1994 {\it \PPCF} \href{http://dx.doi.org/10.1088/0741-3335/36/12/001}{{\bf 36} 1869}
    \bibitem{Dudson.2008} Dudson B \etal 2008 {\it \PPCF} \href{http://dx.doi.org/10.1088/0741-3335/50/12/124012}{{\bf 50} 124012}
    \bibitem{Ayed.2009} {Ben Ayed} N \etal 2009 {\it \PPCF} \href{http://dx.doi.org/10.1088/0741-3335/51/3/035016}{{\bf 51} 035016}
    \bibitem{Fuchert.2014} Fuchert G \etal 2014 {\it \PPCF} \href{http://dx.doi.org/10.1088/0741-3335/56/125001}{{\bf 56} 125001}
    \bibitem{Stober.2000} Stober J \etal 2000 {\it \PPCF} \href{http://dx.doi.org/10.1088/0741-3335/42/5A/324}{{\bf 42} A211}
    \bibitem{Hennequin.2015} Hennequin P \etal 2015 {\it 42nd EPS Conference} 
\end{thebibliography}
\end{document}